\documentclass[sigconf,nonacm]{acmart}
\renewcommand\footnotetextcopyrightpermission[1]{} 
\AtBeginDocument{%
  \providecommand\BibTeX{{%
    \normalfont B\kern-0.5em{\scshape i\kern-0.25em b}\kern-0.8em\TeX}}}

\usepackage{listings}
\definecolor{mygreen}{rgb}{0,0.6,0}
\definecolor{mygray}{rgb}{0.5,0.5,0.5}
\definecolor{mymauve}{RGB}{255,127,80}

\begin{document}

\title{DISCOVER: A Data-driven Interactive System for Comprehensive Observation, Visualization, and ExploRation of Human Behaviour}


\author{Dominik Schiller}
\email{dominik.schiller@uni-a.de}
\orcid{1234-5678-9012}
\affiliation{%
  \institution{University of Augsburg}
  \streetaddress{Universitätsstraße 6a}
  \city{Augsburg}
  \state{Bavaria}
  \country{Germany}
  \postcode{86159}
}

\author{Tobias Hallmen}
\email{tobias.hallmen@uni-a.de}
\orcid{0009-0005-6450-5694}
\affiliation{%
  \institution{University of Augsburg}
  \streetaddress{Universitätsstraße 6a}
  \city{Augsburg}
  \state{Bavaria}
  \country{Germany}
  \postcode{86159}
}

\author{Daksitha Withanage Don}
\email{daksitha.withanage.don@uni-a.de}
\orcid{1234-5678-9012}
\affiliation{%
  \institution{University of Augsburg}
  \streetaddress{Universitätsstraße 6a}
  \city{Augsburg}
  \state{Bavaria}
  \country{Germany}
  \postcode{86159}
}

\author{Elisabeth André}
\email{elisabeth.andre@uni-a.de}
\orcid{1234-5678-9012}
\affiliation{%
  \institution{University of Augsburg}
  \streetaddress{Universitätsstraße 6a}
  \city{Augsburg}
  \state{Bavaria}
  \country{Germany}
  \postcode{86159}
}

\author{Tobias Baur}
\email{tobias.baur@uni-a.de}
\orcid{0000-0002-2797-605X}
\affiliation{%
  \institution{University of Augsburg}
  \streetaddress{Universitätsstraße 6a}
  \city{Augsburg}
  \state{Bavaria}
  \country{Germany}
  \postcode{86159}
}

\renewcommand{\shortauthors}{Schiller, et al.}



\newcommand{\toolname}{\mbox{\textit{DISCOVER }}}

\begin{abstract}
Understanding human behavior is a fundamental goal of social sciences, yet its analysis presents significant challenges.
Conventional methodologies employed for the study of behavior, characterized by labor-intensive data collection processes and intricate analyses, frequently hinder comprehensive exploration due to their time and resource demands.
In response to these challenges, computational models have proven to be promising tools that help researchers analyze large amounts of data by automatically identifying important behavioral indicators, such as social signals.
However, the widespread adoption of such state-of-the-art computational models is impeded by their inherent complexity and the substantial computational resources necessary to run them, thereby constraining accessibility for researchers without technical expertise and adequate equipment.
To address these barriers, we introduce \toolname-- a modular and flexible, yet user-friendly software framework specifically developed to streamline computational-driven data exploration for human behavior analysis.
Our primary objective is to democratize access to advanced computational methodologies, thereby enabling researchers across disciplines to engage in detailed behavioral analysis without the need for extensive technical proficiency.
In this paper, we demonstrate the capabilities of \toolname using four exemplary data exploration workflows that build on each other: Interactive Semantic Content Exploration, Visual Inspection, Aided Annotation, and Multimodal Scene Search.
By illustrating these workflows, we aim to emphasize the versatility and accessibility of \toolname as a comprehensive framework and propose a set of blueprints that can serve as a general starting point for exploratory data analysis.

\end{abstract}




\begin{teaserfigure}
  \includegraphics[width=\textwidth]{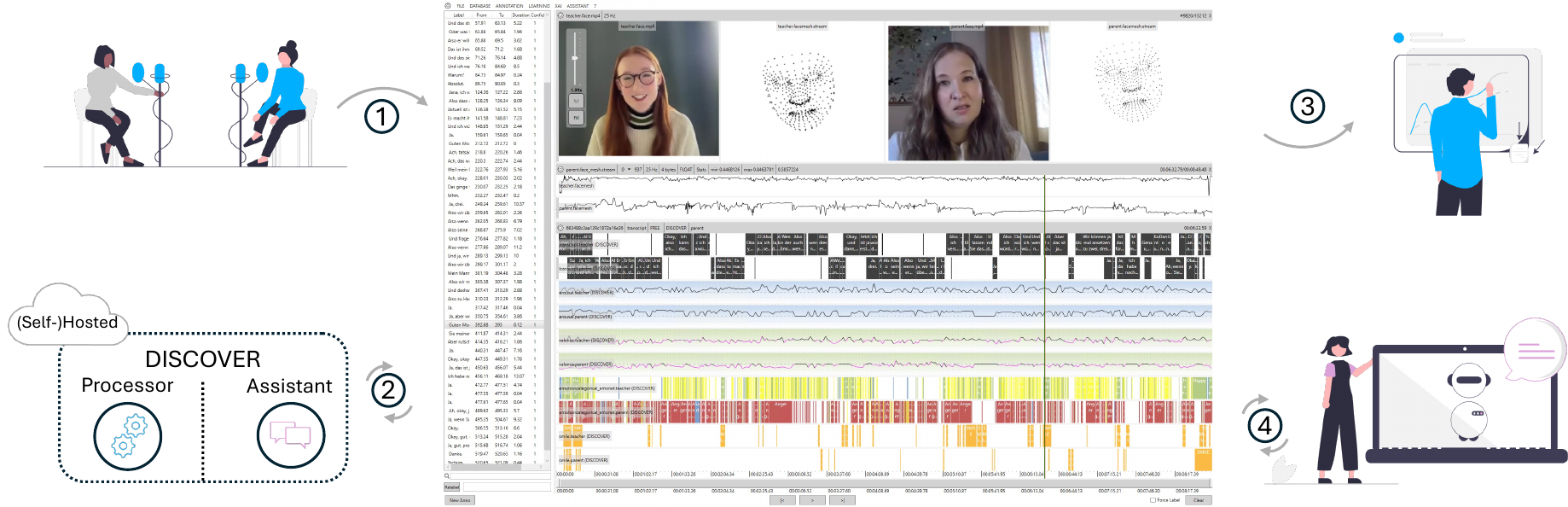}
  \caption{Overview of the \toolname Architecture}
  \Description{Overview of the Nova-Server Architecture}
  \label{fig:teaser}
\end{teaserfigure}


\maketitle
\thispagestyle{empty}

\section{Introduction}

Understanding human behavior is a core pursuit in social sciences, crucial for unraveling the complexities of human interactions and societal dynamics. 
However, analyzing human behavior comes with its fair share of challenges, particularly with traditional methods that involve laborious data collection and intricate analyses, often limiting thorough exploration due to resource demands.

To address these challenges, computational models have emerged as promising alternatives, offering automated identification of key behavioral cues like social signals. 
Yet, their widespread adoption faces hurdles, mainly due to their complexity and the hefty computational resources they require, making them less accessible to researchers without technical expertise or adequate equipment.

In light of these issues, we introduce \toolname \footnote{https://github.com/hcmlab/discover}, a flexible open-source software framework tailored to streamline computational-driven data exploration for human behavior analysis. 
Our primary goal is to democratize access to advanced computational tools, empowering researchers from diverse backgrounds to conduct in-depth behavioral analysis without extensive technical know-how.

Throughout this paper, we showcase \toolname's versatility through four illustrative data exploration workflows.
First, we describe the \textit{interactive semantic content exploration} of transcriptions, achieved through direct integration of large language models.
Second, we illustrate how the integrated social signal processing capabilities of \toolname can aid researchers in the \textit{visual exploration} of data.
Third, we show how \toolname can be used to \textit{aid the annotation} of meaningful behavioral cues.
Lastly, we showcase our framework's capabilities for \textit{multimodal scene search}, enabling users to automate the identification of key scenes within datasets. 

Although all suggested workflows can be applied individually and adapted to user-specific needs, we suggest to consider them as blueprints for starting data exploration with \toolname.

\section{Related Work}




Previous research in the field of human-computer interaction has suggested methods for displaying multimodal characteristics within particular conversational contexts to analyze human behavior. 

Over time, a range of annotation tools concentrating on affective computing and social cues has emerged, offering assistance to users in their efforts.
Prominent examples include \textsc{ELAN} \cite{Wittenburg:2006}, \textsc{ANVIL} \cite{Kipp:2013}, and \textsc{EXMARALDA} \cite{Schmidt:2004}. 
These tools offer layer-based tiers to insert time-anchored labeled segments, that we call \emph{discrete} annotations. 
\emph{Continuous} annotations, on the other hand, allow an observer to track the content of an observed stimulus over time based on a continuous scale. 
One of the first tools that allowed labelers to trace emotional content in real-time on two dimensions (activation and evaluation) was \textsc{FEELTRACE} \cite{Cowie:2000}. 
Its descendant \textsc{GTRACE} (general trace) \cite{Cowie:2012} allows the user to define their own dimensions and scales. 
More recent tools to accomplish continuous descriptions are \textsc{CARMA} (continuous affect rating and media annotation) \cite{Girard:2014} and DARMA (dual axis rating and media annotation) \cite{Girard:2016}.

Recently, these tools have evolved to include automatic calculation of behavioral cues and social signals, eliminating the need for the manual annotation of data.
For example \textsc{Emodash} \cite{ez2020emodash} has been designed to enhance tutors’ retrospective understanding of learners’ emotions, based on facial expressions, within a video-conferencing learning setting.
\textsc{REsCUE} \cite{arakawa2019rescue} helps to aid coaching practitioners to detected unconscious behavior of their clients. To this end, \textsc{REsCUE} uses an unsupervised anomaly detection algorithm to cluster.
\textsc{MultiSense} \cite{zeng2020multisense} can assess the affective state of a person by inferring various indicators from audio-visual input signals. 
The tool focuses on the application within the mental health domain to assess indicators of psychological distress such as depression or post-traumatic stress disorder.
\textsc{MeetingCoach} \cite{samrose2021meetingcoach} is an AI-driven feedback dashboard designed to enhance the effectiveness and inclusively of video-conferencing meetings by providing personalized insights into meeting dynamics, such as engagement summaries of participants or speaking time distribution.
\textsc{MACH} \cite{hoque2013mach} social skills training, particularly focusing on job interview preparation, that analyses non-verbal behavior of an interviewee. 
The \textsc{AffectToolbox} \cite{mertes2024affecttoolbox} provides a software system aimed at aiding researchers in developing affect-sensitive studies and prototypes in affective computing.
It provides accessible functions for analyzing users' affective states via a graphical user interface, including a variety of emotion recognition models for different modalities as well as a unified multimodal assessment.
The \textsc{ConAn} Tool \cite{penzkofer2021conan} has been developed with a focus on group conversation analysis.  
To this end, it automatically analyzes the gaze behavior, body movement, speaker activity, and facial expressions of participants using a single 360\textdegree camera. 

All these visualization-based methods were developed with specific goals and target groups in mind.
As a result, the choice of features to be displayed is usually tailored to this specific use case.
Therefore these solutions suffer from a lack of customizability that prevents users from adapting the features to their individual needs.

The \textsc{Social Signal Interpretation Framework}(SSI) by \citet{wagner2013social} presents an alternative approach, by implementing a modular, multimodal signal processing pipeline, facilitating both online and offline recognition tasks. 
The plug-in system within \textsc{SSI} allows users to develop custom modules and integrate them into the processing pipeline. 
Similar to \textsc{SSI} the \textsc{Opensense} \cite{aberer2010opensense} platform has been designed to facilitate real-time acquisition and recognition of social signals through multiple modalities. 
It also follows a modular pipeline design and builds on Microsoft's Platform for Situated Intelligence \cite{bohus2021platform}, which enables the processing of human behavioral signals and supports various sensor devices and machine learning tools. 
\citet{barz2021multisensor} developed the \textsc{MultiSensorPipeline}, a lightweight and adaptable framework for creating multimodal-multisensor interfaces using real-time sensor data. 
While the framework is conceptually similar to \textsc{SSI} and \textsc{Opensense} it focuses on a concise set of concepts and functionalities that ease the creation and execution of complex processing pipelines.
The process of implementing specific modules is left to the user, making the tool better suited for developing prototypes of custom multimodal-multisensor processing pipelines than for using standard modules to analyze social signals.
In contrast to the pure visual exploration of data, \textsc{Providence} implements a scene search approach to investigate social behavior in multimodal data. 
Hereby, a human analyst can formulate queries containing non-verbal (e.g. nodding, facial expressions), linguistic (e.g. sentiment), or para-linguistic (e.g. speech speed or volume) to search for specific scenes in a human conversation. 
\textsc{Providence} automatically extracts the respective features required by a query and searches for scenes fulfilling the specified conditions in the data. 

Although the approaches presented have their respective advantages and disadvantages, there are two common shortcomings: Efficient use of computing resources and a compromise between flexibility and complexity.
Except for \textsc{Providence}, all the tools introduced are solely intended for operation on local machines. 
This leads to inefficient use in multi-user scenarios, in which users either have to take turns on a local machine or each user requires his own workstation that is capable to run such tools.
Considering hardware demands and energy consumption of state-of-the-art machine learning models, this issue extends beyond financial concerns to ecological ones.
Further, it becomes evident that off-the-shelf solutions have constraints in their applicability, whereas more adaptable approaches necessitate greater technical proficiency, thereby posing a barrier to entry.

\section{Framework}\label{sec:ecosystem}

\begin{figure}[h]
    \centering
    \includegraphics[width=\linewidth]{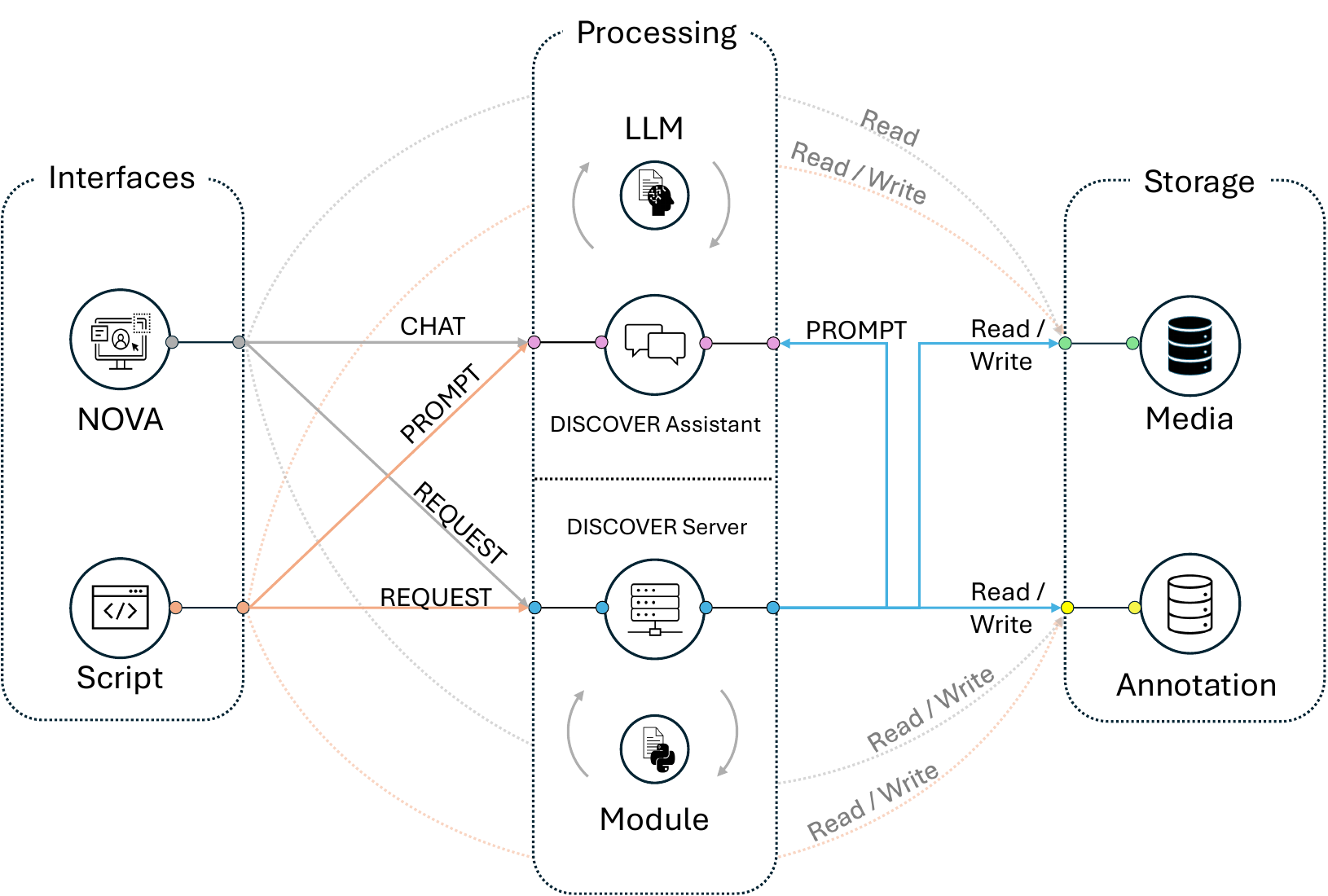}
    \caption{Overview of the system components for the \toolname eco system}
    \label{fig:system_overview}
\end{figure}

The architecture of \toolname follows a modular design, which facilitates flexibility, scalability, and ease of maintenance. 
At its core, the framework comprises several key components, each serving a distinct purpose in enabling laymen to perform data exploration tasks efficiently. 
An overview of the system architecture is shown in Figure \ref{fig:system_overview}.
Below, we provide a textual skeleton outlining the main components and their functionalities.

\subsection{User Interface}
Each element in the \toolname framework utilizes APIs to communicate over the network, granting users with programming skills significant flexibility by allowing access via scripts, irrespective of the programming language employed.
In order to make \toolname more user-friendly for individuals without technical expertise, we've incorporated its API into the open-source tool NOVA \cite{Heimerl2019}, to serve as the graphical user interface. 
NOVA aims to enhance the standard annotation process with the latest developments from contemporary research fields such as Cooperative Machine Learning and eXplainable Artificial Intelligence by giving annotators easy access to automated model training and prediction functionalities, as well as sophisticated explanation algorithms via its user interface.

\begin{figure}[ht]
\includegraphics[width=0.48\textwidth]{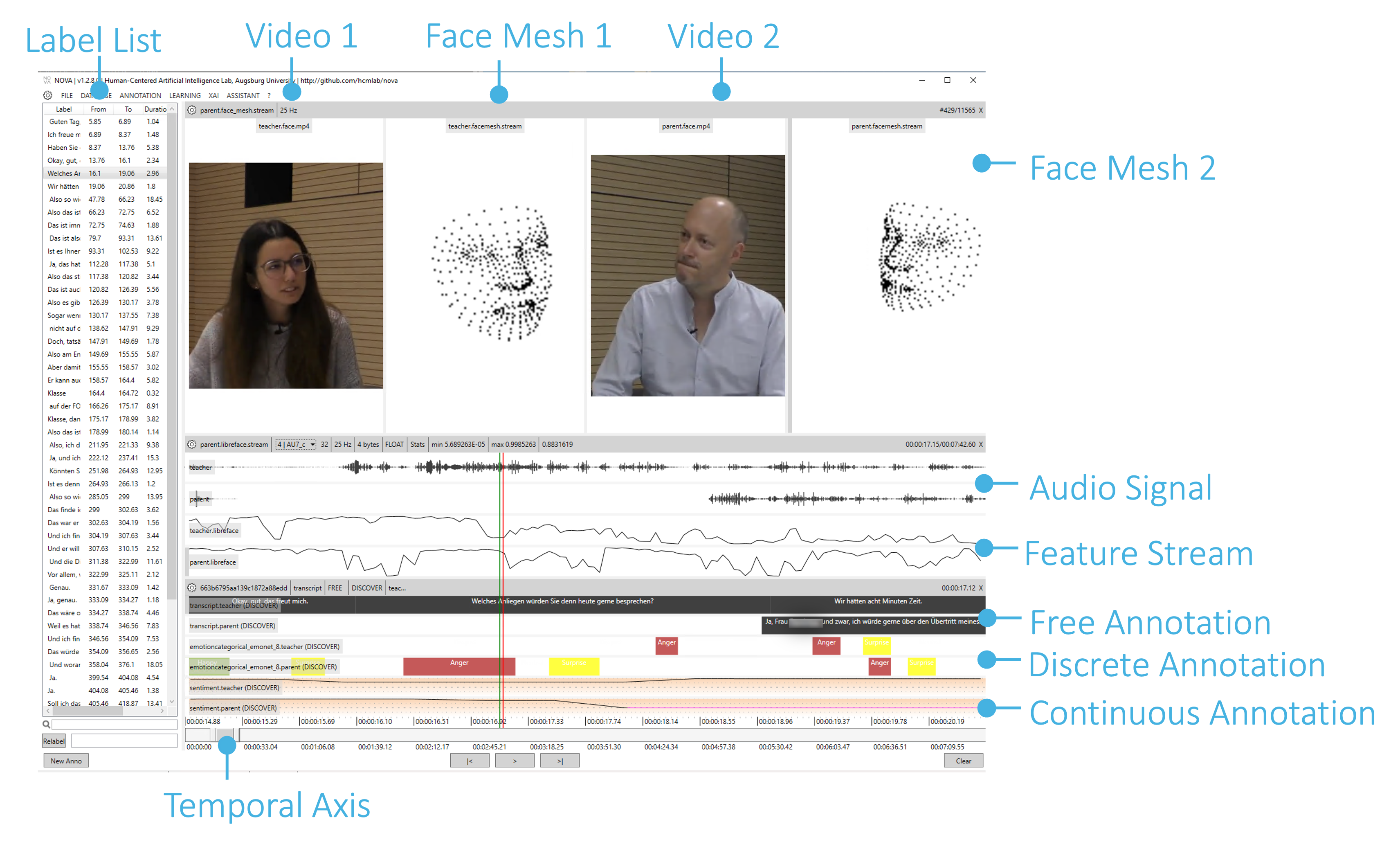}
\caption{\textsc{NOVA} allows to visualise various media and signal types and supports different annotation schemes. From top downwards: upper-body videos along with face tracking, audio streams of two persons during an interaction, and activation of the action units. 
In the lower part, free-value, discrete, and continuous annotation tiers are displayed.
}
\label{fig:nova:overall}
\end{figure}

The NOVA user interface has been designed with a special focus on the annotation of long and continuous recordings involving multiple modalities and subjects.
A screenshot of a loaded recording session is shown in Figure \ref{fig:nova:overall}. 
On the top, several media tracks are visualized and ready for playback. 
Note that the number of tracks that can be displayed at the same time is not limited and various types of signals (video, audio, facial features, skeleton, depth images, etc.) are supported. 
In the lower part, we see multiple annotation tracks of different types (discrete, continuous, and transcriptions) describing the visualized content. 

NOVA provides several functions to process the annotations created by multiple human or machine annotators. 
For instance, statistical measures such as Cronbach's $\alpha$, Pearson's correlation coefficient, Spearman's correlation coefficient or Cohen's $\kappa$ can be applied to identify inter-rater agreement.

\subsection{Annotation Database}

To support a collaborative annotation process, \toolname maintains a database back-end, which allows users to load and save annotations from and to a MongoDB\footnote{\url{https://www.mongodb.com/}} database running on a central server. 
This gives annotators the possibility to immediately commit changes and follow the annotation progress of others. 
Besides human annotators, a database may also be visited by one or more ``machine users''. 
Just like a human operator, they can create and access annotations. 
Hence, the database also functions as a mediator between human and machine. 
\toolname provides instruments to create and populate a database from scratch. 
At any time new annotators, schemes, and additional sessions can be added. 

\subsection{Media File Storage}

\toolname uses a data storage component that follows the structure of the open-source cloud hosting framework Nextcloud\footnote{\url{https://nextcloud.com/de/}}. 
Data can be hosted on a local drive and be shared on demand with people who have been granted access to the respective Database.
Both the NOVA user interface, as well as the processing backend can access these files for visualization and processing respectively.

\subsection{Processing Server}
The processing server acts as the computational engine powering data analysis tasks within our framework. 
It implements a lightweight web server that interacts with the annotation and data storage components to extract meaningful information from the data, that can subsequently be visualized in NOVA.
The server-based architecture of \toolname provides key advantages concerning flexibility, accessibility, and resource efficiency. 
First of all, exposing a REST API makes the server not only accessible from the NOVA user interface but also via scripts. 
This results in a low entry barrier for not coding affine research, while maintaining the ability to include the processing functionality in custom scripts. 
Second, the server-based approach enables multiple users to process data in a central place to ensure optimal usage of computational resources. 
The standardized input and output formats of the processing modules also enable the sharing and further usage of results.

\subsection{Assistant}
Recently large language models (LLM), like Chat-GPT\footnote{\url{https://chatgpt.com/}} or LLama \cite{touvron2023llama} have demonstrated remarkable capabilities for textual analysis tasks like text summarization \cite{luo2023chatgpt, liu2024exploring, soni2023comparing}, sentiment analysis \cite{fatouros2023transforming, zhang2023sentiment} or argument mining \cite{irani2024wiba, pojoni2023argument}.
To incorporate such capabilities into \toolname, we supplement the processor component with an assistant.
Fundamentally the assistant is a lightweight web server that aims to integrate numerous LLMs with the rest of the \toolname infrastructure. 
To this end, the assistant is exposing a unified API that reroutes requests to either an external service provider (e.g. OpenAI), a self-hosted large language model (e.g. via Ollama \footnote{\url{https://ollama.com/}}.
Through seamless integration with the UI, users can interact with AI assistants via a chat interface to analyze textual data, such as dialogue transcripts. 
Depending on the performance of a model for specific tasks and privacy requirements a user can switch between available services dynamically during the exploration process. 

\section{Modules} \label{sec:modules}
\toolname relies on exchangeable modules to infer behavioral indicators from recorded data. 
Each of these modules can be understood as a configurable building block, consisting of predefined inputs and outputs with module-specific options. 
\toolname is fully extendable with custom modules. 
However, to keep the entrance barrier low and provide value for a non-coding affine target group we also provide a number of ready-to-use processing modules. 
The following section provides an overview of the currently integrated modules.

\subsection{Face}
\paragraph{Bounding Box Detector}
The automatic derivation of behavioral indicators from a human face usually requires the localization of the facial area in an image or video.
To this end, we rely on the BlazeFace model proposed by \cite{bazarevsky2019blazeface} et al. 
The BlazeFace model is a lightweight face detection model that has been developed to run on mobile devices and thus requires only a minimum of computational resources to achieve super-realtime performance.  
\paragraph{Landmarks and Meshes}
For the further processing of the localized image, it's it is a common procedure to align facial images based on localized landmarks \cite{chang2023libreface}. 
Facial image alignment involves geometric transformations like translation, rotation, and scaling to convert the input face image into a standardized form.
To this end, we employ the face mesh model by \citet{kartynnik2019real}, which infers an approximate 3D mesh representation of a human face from a single camera.
\paragraph{Action Units}
Facial action units originate from an anatomical examination of the face and can be categorized according to the Facial Action Coding System (FACS) outlined by \citet{ekman1976measuring}.
For automatic action unit detection and intensity estimation, we integrated the LibreFace~\cite{chang2023libreface} framework which achieves state-of-the-art performance in both tasks while improving inference times over other methods.
\paragraph{Facial Expression}
Facial expression analysis involves automatically detecting subtle movements in facial muscles and identifying typical facial displays. 
The recognition of these expressions yields valuable insights into users' social and emotional states.
To facilitate robust facial expression prediction we integrated multiple models into \toolname Emonet~\cite{toisoul2021estimation}, Relevance-based data masking~\cite{schiller2020relevance}, LibreFace~\cite{chang2023libreface}.

\subsection{Voice}
When analyzing human speech to gain behavioral insights, one needs to distinguish between the verbal and vocal components of speech.
Verbal refers to communication that is expressed in words or language. 
It involves the use of language to convey ideas, thoughts, or information. 
Vocal on the other hand refers to the sounds produced by the voice or the act of speaking. 
In the context of communication, "vocal" can refer to the tone, pitch, volume, and other qualities of speech. 

\subsubsection{Verbal}
A prerequisite to the analysis of the verbal content of spoken language is the conversion from speech to text (STT). 
STT-Systems have been an active area of research for decades.
For the implementation of our STT module we rely on \textsc{WhisperX}~\cite{bain2022whisperx}, an adaptation of the \textsc{Whisper} Model \citet{radford2023robust}, that provides improved timestamp accuracy, support for longer audio sequences, and faster transcription performance.

\paragraph{Speaker Diarization}
The available datasets are typically recorded with the focus of manual human analysis rather than automatic processing.
A common example of this is the recording of a single audio signal for several speakers.
While it is a mostly trivial task for a human listener to distinguish between the voices of speakers and map the content of the spoken language to the respective person, this information gets lost during the STT process. 
To account for this loss of information \toolname implements a speaker diarisation module, which maps segments of a common dialogue transcript to individual speakers.
To this end we rely on \textsc{Pyannote} \cite{Bredin2020, Bredin2021} to cluster voiced segments in the audio signal. 
We then use an oracle approach to assign those clusters to individual speakers, by providing reference speaking terms within the audio signal.

\paragraph{Sentiment Analysis}
Sentiment analysis is the process of computationally determining the emotional tone behind a piece of text, whether it's positive, negative, or neutral. 
It can be a valuable tool for analyzing human behavior, as it can provide a deeper understanding of individuals' emotions and opinions.
To enable the automatic prediction of sentiment, \toolname currently integrates two approaches. 
A multi-lingual \cite{barbieri-etal-2022-xlm} ) and a German language specific \cite{guhr2020germansentiment} model.

\subsubsection{Vocal}
Speech emotion recognition (SER) refers to the task of automatically detecting and interpreting emotions conveyed through speech. 
It involves analyzing various acoustic features, such as pitch, intensity, and rhythm, to infer the underlying emotional state of the speaker.
\toolname integrates a pre trained model, proposed by \citet{wagner2023dawn} to automatically detect valence, arousal, and dominance values from a human voice.

\subsection{Multimodal Feature Extraction}
Besides the above-mentioned modules, which directly provide insights about important indicators for human behavior to an analyst, \toolname also implements modality-specific feature extraction modules, that can be used to train custom detection models.

\paragraph{Video}
For the video modality we use the DinoV2 pretrained vision transformer models \cite{oquab2023dinov2, darcet2023vitneedreg} to extract features.
Those models are pretrained in a self-supervised manner on a large dataset of 142 million images.
As a result, DINOv2 models have demonstrated robust performance beyond training data, delivering usable general-purpose features without the need for fine-tuning.

\paragraph{Audio}
For the audio modality we rely on a pretrained w2v-BERT 2.0 encoder \cite{barrault2023seamless}. 
Similar to the DinoV2 model, this model was trained unsupervised on a large data set of 4.5 million hours of audio, and demonstrates excellent performance for a variety of downstream tasks like speech to text, or expressive speech to speech translation.
However, it is recommended to fine-tune the w2v-BERT 2.0 model before using it for a downstream task. 
Since this might not be feasible for technical unsavvy users we also integrated the openSMILE library \cite{eyben2010opensmile}, which extracts various handcrafted feature sets for the audio domain.
Specifically, the GeMaps feature set \cite{eyben2015geneva} has been developed for general voice research and affective computing tasks and provides a good starting point for any speech-related classification task.

\paragraph{Text}
When it comes to extracting features from text representations, the language of the text is a necessary consideration. 
Since it is a key aspect of our framework to be employable in versatile scenarios across multiple languages, \toolname integrates a multilingual textual feature extraction using the XLMroBERTa (XLM-R) model by \citet{unsupervised2019conneau}
This model consists of a transformer-based architecture, trained on vast amounts of multilingual data crawled from the internet.
In their experiments, the authors analyzed the capabilities of XLM-R for several tasks, including name entity recognition, cross-lingual question answering, paraphrasing, and sentiment analysis.
The reported results indicate that the model performs close to or even better than comparable monolingual models for languages where vast training resources are available. 
Furthermore, the model showed substantial improvements over other state-of-the-art models on low-resource languages across all tasks.

\section{Interactive Data Exploration}

\begin{figure}
    \centering
    \includegraphics[width=0.48\textwidth]{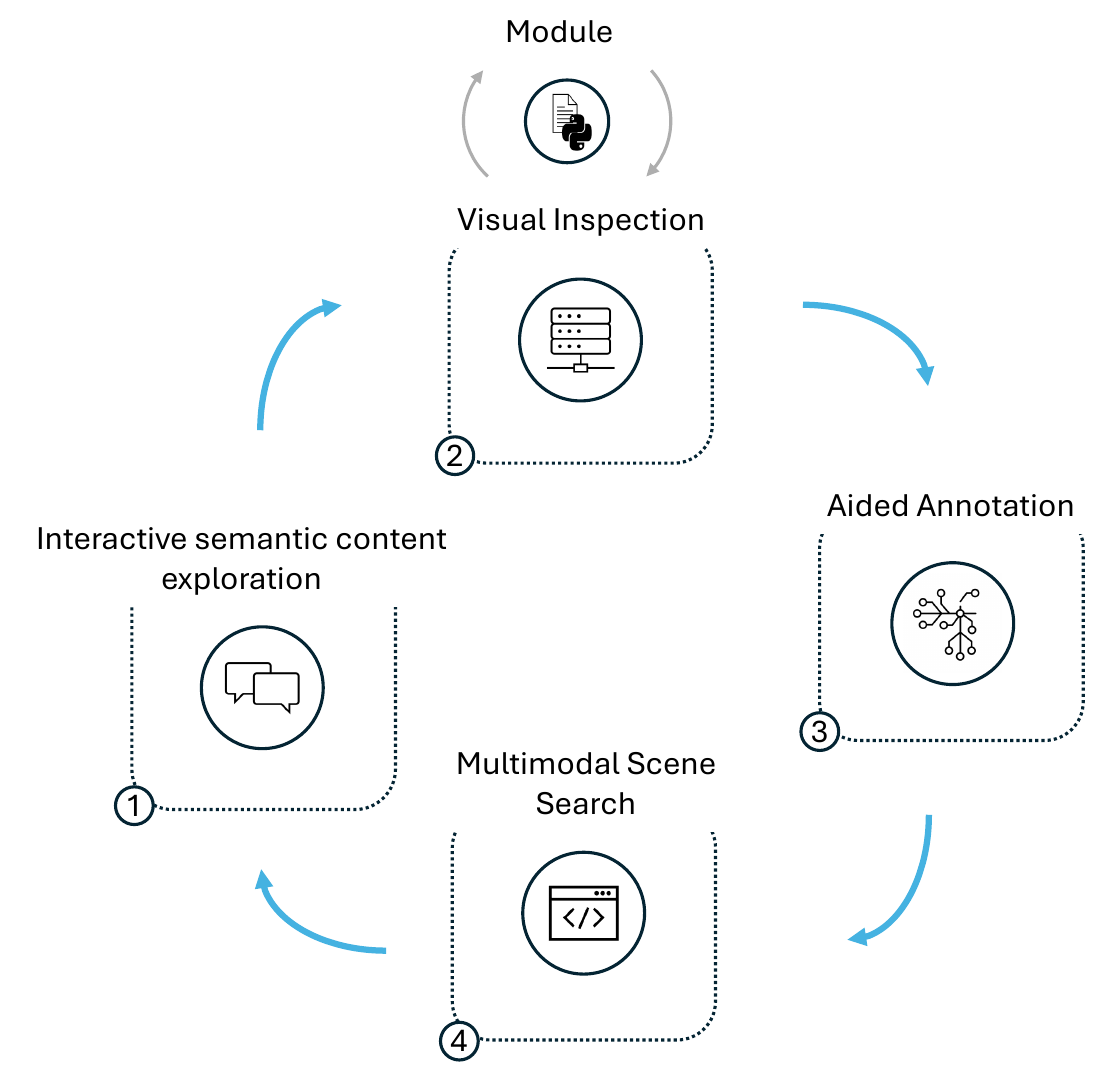}
    \caption{Schematic representation of the exploratory data analysis workflow. 1. Explore the content of a conversation using the \toolname assistant. 2. Compute and visualize behavioral cues. 3. Annotate additional indicators with the help of cooperative machine learning. 4. Identify scenes that consist of an interplay of several indicators. Every step in the workflow can be repeated and adapted as necessary.}
  \label{fig:workflow}
\end{figure}

\subsection{Semantic Content Exploration} 
\label{sec:isce}
\begin{figure}
    \centering
    \includegraphics[width=\linewidth]{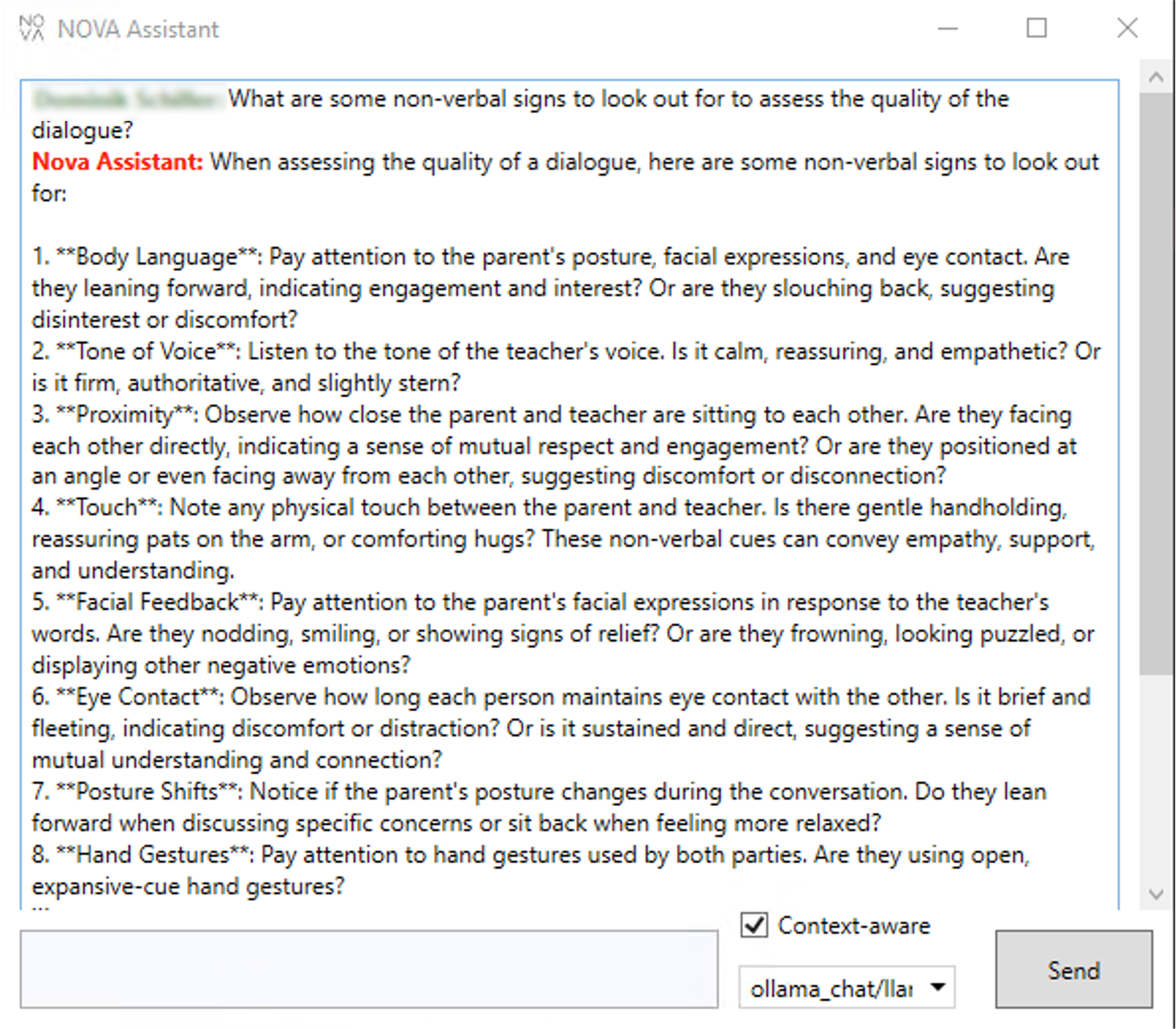}
    \caption{A user engages with the assistant to explore the semantics of a conversation interactively.}
  \label{fig:assistant}
\end{figure}

\begin{figure*}
    \centering
    \includegraphics[width=0.98\textwidth]{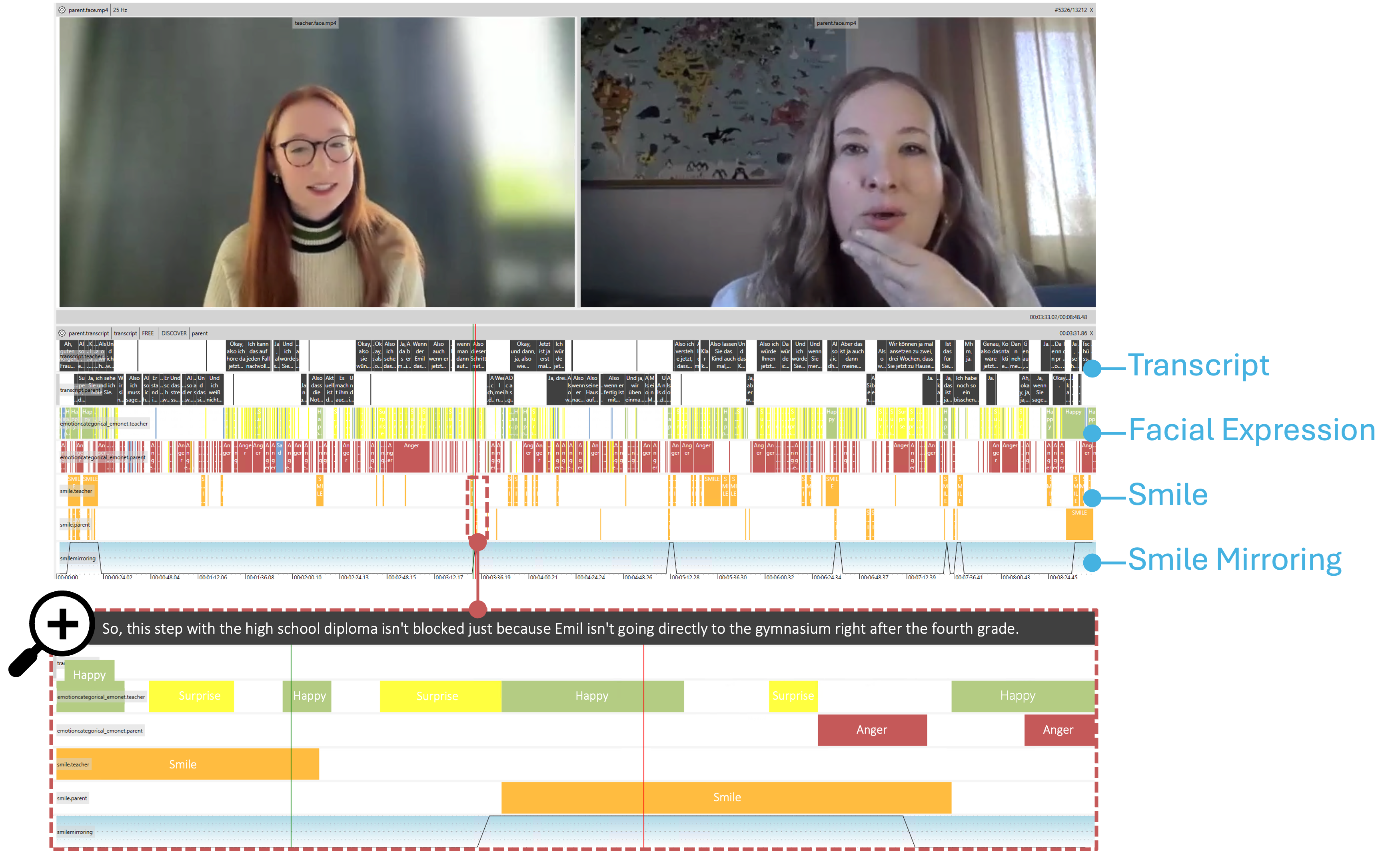}
    \caption{Visual data inspection with \textsc{NOVA} and \toolname. 
    The session overview provides a comprehensive summary of the extracted behavior indicators across the entire session. Zooming in at the bottom enables a detailed analysis of identified scenes of interest.}
  \label{fig:workflow_example}
\end{figure*}

In the following section, we demonstrate the capabilities of \toolname using four exemplary workflows to examine unseen data and gather new insights. 
While each of our showcases is building upon the results of the previous one, every workflow can also be carried out independently of the others.
The complete iterative data exploration pipeline is depicted in Figure \ref{fig:workflow}.

We demonstrate how our workflow operates by utilizing recordings of interactions between teachers and parents. 
These videos are captured to evaluate the communication abilities of aspiring teachers in consultative situations and offer them constructive feedback. 
Within this context, the teacher is a trainee, while the role of the parent is portrayed by an actor.
The subject of the discussion revolves around the child of the parent, who is facing challenges in school.
Since the original discussion is in German a translated version can be found in Appendix A1.
The following is an example workflow of the use case for data analysis from an analyst's point of view.

As a first step towards finding indicators of communicative quality in the recorded interactions, a user wants to get familiar with the data and the task. 
To facilitate those tasks, \toolname assistant enables the interactive exploration of the semantic content of the dialogue.
To begin using the assistant a user first utilizes the WhisperX processing module to create a temporal-aligned transcript for each speaker from the recorded audio signal (see Appendix A1).

After loading the transcription data into \textsc{NOVA} the user clicks on the Assistant tab and a chat window opens that establishes a connection with the assistant (see Figure \ref{fig:assistant}). 
By checking the "Context-Aware" checkbox in the bottom \textsc{NOVA} the transcript that has been loaded into \textsc{NOVA} before will be automatically sent to the assistant accompanying each message.
That way a user can directly ask the assistant any questions regarding the semantic content of the dialogue. 
Firstly the user requests a summary of the interaction.
The assistant replies with a concise summary of the transcript, providing the user with information about the general setting and topic of the dialogue, the roles of the interlocutors, and the course of the conversation.
As the user has gained those insights about the data, the next step is to gather more information on relevant behavioral aspects to look out for.
To this end, the user asks the assistant directly about important criteria to assess the quality of communication in parent-teacher conferences. 
The assistant answers by providing ten indicators for the assessment of dialogue quality, like \textit{Positive outcome} or \textit{Collaborative approach}.
Each point is accompanied by a short description to clarify the meaning. 
From the previous summary, it is already clear that the teacher and parent are working together and the outcome of the dialog is positive as both parties agree on how they want to proceed in the future to support the child's learning.
The user then continues to ask the assistant to evaluate the transcript concerning the now-identified indicators to get deeper insights beyond the summary.
In return, the assistant provides further information about indicators like \textit{Empathy} and \textit{Active Listening}, based on the full transcript. 
Finally, the user also wants to know about non-verbal indicators for communication quality, to which the assistant again provides a list of key points to look out for like \textit{Facial Expressions} / \textit{Smiles}, \textit{Gaze} or \textit{Vocal Inflections and Tone}.
Appendix A2 shows the complete dialogue between the user and the assistant.

\subsection{ Aided Annotation }
\label{sec:cml}
The processing modules described in Section \ref{sec:modules} are a core aspect of \toolname. 
However, they largely depend on the availability of pre-trained models for processing.  
Depending on the use case there might be no fitting model available for the task that the user is looking for. 
To alleviate this problem, \toolname provides support for feature extraction, that can be used to train custom models directly from within the NOVA user interface. 
To this end, \textsc{NOVA} implements a cooperative machine learning workflow that consists of the following 5 steps:
Initially, the model undergoes training (Step 1) and is then utilized to predict unseen data (Step 2). 
Following this, an active learning module determines which portions of the prediction necessitate manual review by human annotators (Step 3). 
Subsequently, those labels are then reviewed by a human and corrected if necessary.
Finally, the initial model is retrained using the updated labeled data (Step 5). 
This iterative process continues until all data is annotated. 
By actively integrating human expertise into the learning process, we enable interactive guidance and enhancement of automatic predictions. 
Following the suggestions of the \toolname Assistant, the user trains a new model that is able to detect smiles, based on the extracted face mesh data.
Before continuing the next step the model is used to detect all instances of smiles for the teacher as well as the parent.

\begin{figure}[hbtp]
    \caption{Sample code demonstrating how to create an annotation for detecting smile mirroring.}
    \label{fig:code}
    \begin{lstlisting}[language=Python,frame=single, basicstyle=\tiny,]
        # Inputs
        smile_teacher = {
            "src": "db:annotation:Discrete",
            "type": "input",
            "id": "smile_teacher",
            "scheme": "smile",
            "role": "teacher",
            "annotator": "DISCOVER",
        }
        
        smile_parent = {
            "src": "db:annotation:Discrete",
            "type": "input",
            "id": "smile_parent",
            "scheme": "smile",
            "role": "parent",
            "annotator": "DISCOVER",
        }
        
        ds_iterator = DatasetIterator(
            dataset="dataset_name",
            session_names="session_name",
            data_description=[smile_teacher, smile_parent],
            frame_size=40,
            left_context=1960,
            source_context=<DB_CREDENTIALS>,
        )
        ds_iterator.load()
        
        # Processing
        output = []
        for sample in ds_iterator:
            mirror = ((sample["smile_teacher"][0] != 0) and 
                      (sample["smile_parent"][0] != 0))
            score = int(mirror)
            confidence = 1
            output.append((score, confidence))
        
        # Creating the annotation
        annotation_scheme = ContinuousAnnotationScheme(
            name="smile_mirroring", 
            sample_rate = 25, 
            min_val=0, 
            max_val=1
            )
        annotation = ContinuousAnnotation(
            scheme=annotation_scheme, 
            data=np.asarray(output, dtype=annotation_scheme.label_dtype)
            )
        fh = FileHandler()
        fh.save(data=annotation, fp='./smile_mirroring.annotation')
    \end{lstlisting}
\end{figure}


\subsection{Visual Inspection}
\label{sec:gvi}
During this stage, the user visually examines the session by navigating through sessions and their timeline. 
The objective at this stage is to identify patterns and formulate hypotheses regarding their manifestation, aiming to extract further insights. 
To confirm or question these hypotheses, users can select a processing module (see Section \ref{sec:modules}) directly from the \textsc{NOVA} User interface and start an extraction job on the processing server. 
Once the processing is done, the results can be directly visualized in the UI. 
This iterative process can be repeated as often as necessary for different modules.
Since the results of the previous modules are stored either in the annotation database or on the media storage, the user can reuse them at any time, without recomputing them. 
To provide a clearer illustration of this process, we pick up on the previous example.
Building upon the results of the semantic content exploration described in \ref{sec:isce} the next steps are analyzing the non-verbal cues from the audio and the video signal and finding additional semantic indicators to assess the quality of communication based on the transcript. 

To continue the exploration, the user predicts the facial expressions of both the teacher and the parent using the EmoNet module. 
Upon loading the model's predictions into NOVA. 
As depicted in Figure \ref{fig:workflow_example} it becomes apparent that the teacher's facial expressions predominantly oscillate between "happy" and "surprised," whereas the parent's expressions tend to skew towards anger or sadness, although it's noted that anger may not always be accurate as the parent might simply be displaying a serious expression.
In addition, the user now loads the smile annotation generated in the previous step (see Section \ref{sec:cml}).

Smiling occurs notably more often at the beginning and end of the conversation for both roles, with the teacher exhibiting smiles more frequently throughout the conversation.
As the increased number of smiles at the beginning and end can likely be attributed to formal politeness the user first focuses on the middle sections of the conversation. 
Especially visually identifying, mirroring behavior where the smile of one interlocutor is mirrored by the other one, provides interesting insights.
For example, there is a notable scene in which the parents' smiles mirror the teacher's smile, a moment in which the parents receive new and helpful information, indicating active participation in the dialog.

This observation underscores the significance of integrating visual and verbal cues to capture the dynamics of communication and emotional expression within the interaction.

\subsection { Scene Search }
As evidenced, the exploration of conversational scenes through the analysis of participants' multimodal behavior presents considerable promise for investigating social dynamics.
The process of visual inspection provides an effective method of identifying the constellation of social cues inherent in a specific scene.
However, it is of limited use when it comes to retrieving all instances of such scenes in a recording session. 
First of all, it's easy to overlook certain scenes when scrolling through the timeline of longer recording sessions.
Second, the analysis of social cue constellations that require precise assessment of either timings (e.g. the immediate facial expression to an event) or values (e.g. the degree to which an action unit is activated), can not be efficiently performed this way. 
Lastly, it is easy to see how the reliable identification of patterns that consist of the interplay of multiple clues can quickly become overwhelming for a human analyst. 
To take all these aspects into account, \toolname provides an API that allows the user to define a set of rules that are used to automatically identify scenes based on previously defined conditions. 
Building upon the results of the visual inspection workflow (see Section \ref{sec:gvi}), a user can now follow a blueprint script to download the smile detection annotations for both roles from the database, move a sliding window over the annotations and create a new annotation based on a predefined mirroring condition (see Figure \ref{fig:code}).

\section{Discussion}
\paragraph{Feedback} To obtain initial feedback, we presented \toolname and the proposed workflow to an AI communication researcher who is involved in the interview and feedback process for the prospective teachers in the analyzed video recordings.
In his comments he stated that the potential speed-up of data exploration through the proposed workflow is beneficial for the social science community: \textit{"Basically you are making videos searchable. You can search for specific events as you search for a word in a transcript. This functionality greatly improves the analysis of large volumes of data."}

Further, he mentions that he considered \toolname to be especially useful for inductive analytic strategies, where a researcher reads through the data and allows new concepts to emerge:
\textit{"I can see how this would be useful for grounded theory approaches. In this process, researchers explore data iteratively without any previous assumptions about the findings to come up with new hypotheses and theories." }

On the other hand, he also mentioned how skeptical he is about using the tool to aid deductive strategies where a researcher applies an existing theory on new data. 
He stated that wrong or missing predictions of the automatically inferred cues could have a crucial impact on the results: 
\textit{"When I analyzed the transcripts generated with Whisper, I was looking for filler and backchannels. 
It took me a while to realize, that most of those short sequences are filtered out during the transcription process. 
I needed to add them again manually to be able to use the transcription."}

Lastly, he also suggested how \toolname could be deployed as an interactive feedback tool in the future: 
\textit{"This visualization could also be interesting for the students when providing feedback about their communication skills. It might be interesting to even let them explore their own conversations interactively."}

\paragraph{Limitations and Prospects}
While \toolname boasts a modular architecture, it's worth noting that certain modules may not yield perfect results, potentially introducing inaccuracies or limitations, especially in scenarios requiring precise results. 
Additionally, the absence of certain modules within \toolname could restrict the scope and comprehensiveness of analysis, given the varied nature of human behavior examination. 
On the other hand, the server-based architecture and modular structure of \toolname present notable advantages, facilitating the seamless exchange of processing components and enabling users to easily incorporate new modules or tailor modules to their specific needs. 
This adaptability empowers researchers to customize the tool according to their preferences and requirements. 
Furthermore, \toolname excels in fostering collaboration among researchers by facilitating the sharing of processed results and computing resources, thereby promoting efficient collaboration and nurturing a sense of community within the research domain. 
This collaborative aspect enhances the potential for collective insights and discoveries.

Moreover, \toolname introduces innovative workflow components such as the assistant and cooperative machine learning functionalities, which enhance the efficiency and effectiveness of the analysis process, ultimately yielding more robust and insightful results. 
Despite its weaknesses, including potential module imperfections and missing functionalities, \toolname's strengths in its modular architecture, collaborative features, and unique workflow components position it as a promising tool for democratizing access to advanced computational methodologies in human behavior analysis.

\section{Conclusion}
We introduced \toolname, an open-source tool designed for human behaviour analysis utilizing state-of-the art machine learning models for feature extraction. 
Our framework offers a user-friendly interface through NOVA, streamlining the process and diminishing the necessity for laborious video and audio annotation. 

\toolname can be used to extract various helpful indicators to analyze human behavior such as transcription, facial expression, or emotions.
Furthermore, we presented a prototypical workflow for the exploratory analysis of human behavior in new data. 
In the future, we plan to further enhance the predictive capabilities of \toolname by integrating new processing modules and improving the workflow.

\bibliographystyle{ACM-Reference-Format}
\bibliography{main}


\end{document}


\appendix

\section{Supplementary Material}

\subsection{Transcript}
The entire conversation dialogue utilized in the interactive data exploration workflow is presented here. The transcript has been anonymized and translated into English. No corrections have been made to the grammatical or lexical errors introduced by the transcription module. The segmentation into speaking segments remains unchanged from the output provided by the transcription module.
The comments in square brackets were added by the author to facilitate a better understanding.

\begin{dialogue}
\speak{TEACHER}	Ah, good morning, Mrs. ***.
\speak{PARENT}	Good morning.
\speak{PARENT}	It's great that it works.
\speak{TEACHER}	So, first of all, the technology works. [The conversation is held via Zoom]
\speak{TEACHER}	Can you hear me well?
\speak{PARENT}	Yes, I see you and I hear you.
\speak{TEACHER}	Do you see me well?
\speak{TEACHER}	Does everything work?
\speak{TEACHER}	Okay, super.
\speak{TEACHER}	Then I would like to talk to you right away.
\speak{TEACHER}	We now have roughly five to eight minutes to talk about your request.
\speak{TEACHER}	And I would ask you first of all, why did you want to talk to me today?
\speak{PARENT}	We are very, very worried about the e-mail [Emil] right now.
\speak{PARENT}	I have to say that since he has been in your class in his fourth year, we no longer like his grades.
\speak{TEACHER}	Mhm.
\speak{TEACHER}	Mhm.
\speak{PARENT}	I thought he's done a great job so far.
\speak{PARENT}	He was always kind of on a solid two and now he's kind of slipping this school year.
\speak{PARENT}	And it's always such a close call.
\speak{PARENT}	It always sloshes so close to the two.
\speak{PARENT}	And that really stresses us out because he's supposed to go to grammar school.
\speak{PARENT}	Or what means suppoeds to... he wants that too.
\speak{PARENT}	So he wants to do his A-levels himself and then go to university.
\speak{PARENT}	That's really important to him.
\speak{PARENT}	And it doesn't look like it's going to work out with the grammar school.
\speak{PARENT}	And I don't know what we can do or what you can do, because it's only since this school year, since he's been with you, that things have stopped working.
\speak{TEACHER}	Okay.
\speak{PARENT}	Why?
\speak{TEACHER}	 Okay, so I can hear that you're very worried about Emil, especially now with regard to the transfer, because we're already in February and the transfer is due in May.
\speak{PARENT}	Absolut.
\speak{TEACHER}	I can definitely understand that and do I understand correctly that Emil was actually always quite good in third grade and his grades have only now slipped down to a three and a four in fourth grade, right?
\speak{TEACHER}	Yes.
\speak{TEACHER}	Okay.
\speak{TEACHER}	Yes, as I said, I understand that you're worried.
\speak{TEACHER}	And I would now like to ask you what you would like for Emil now, especially with regard to the transition?
\speak{TEACHER}	What do you want for Emil?
\speak{PARENT}	 Jana, I hope he makes the transition to grammar school.
\speak{PARENT}	 So that the grades are good, that I have my happy Emil again, who comes home and is somehow doing well and where the school day went well.
\speak{TEACHER}	Okay.
\speak{PARENT}	He's also kind of stressed at the moment and when we talk about school, it's so tense.
\speak{PARENT}	He also doesn't enjoy it as much because he doesn't have the success he wants for himself.
\speak{PARENT}	And I hope that he is doing well and that the grades are right.
\speak{TEACHER}	 Okay, so they just want him to be a bit less stressed when he transfers to grammar school.
\speak{PARENT}	Ja.
\speak{TEACHER}	That's the wish.
\speak{TEACHER}	Okay, so I would like to start by describing my situation.
\speak{TEACHER}	Of course, I can also see that Emil's performance is rather mediocre at the moment, and that he's not doing so well.
\speak{TEACHER}	 Yes, it's a bit difficult with the GÜMI at the moment, with the editing.
\speak{TEACHER}	But I can put your mind at rest for now.
\speak{TEACHER}	If Emil really doesn't make it to grammar school at the end of May, there is always the option of transferring to grammar school at the end.
\speak{TEACHER}	So even if he really doesn't make it with this average of 2.33, you don't have to worry that the Abitur and everything will be ruined straight away, because it's not like that.
\speak{TEACHER}	So you can still…
\speak{TEACHER}	once you're at secondary school, if you have a good average of 2.0, for example, then you can always switch back to grammar school.
\speak{TEACHER}	So this step with the Abitur is not blocked just because Emil doesn't go straight to grammar school after the fourth grade.
\speak{PARENT}	Oh, actually, okay.
\speak{PARENT}	Oh, we weren't really aware of that, okay.
\speak{PARENT}	Because my husband and I both went down that route, somehow through grammar school and on to A-levels.
\speak{PARENT}	Oh, okay.
\speak{PARENT}	Theoretically, that would still be possible.
\speak{TEACHER}	 Okay, and then, yes, as I said, I can understand that it's a bit frightening at first if it doesn't work out straight away, but we still have time.
\speak{PARENT}	Mhm.
\speak{TEACHER}	It's only February now and we still have a few months until the transition and maybe things will improve then.
\speak{PARENT}	Yes, three.
\speak{TEACHER}	I'd like to know what you do with Emil at home to maybe support him a little, right?
\speak{PARENT}	So we practise a lot.
\speak{TEACHER}	Mhm.
\speak{TEACHER}	Mhm.
\speak{PARENT}	So when he comes home, we have lunch and then he actually sits down at the driving table and does his things.
\speak{TEACHER}	Mhm.
\speak{PARENT}	That means his homework and I also check in every now and then if he needs help somewhere or wants to make sure that everything is working properly.
\speak{PARENT}	And then ask him a lot of questions.
\speak{PARENT}	So when he's finished with his homework, we always look at the HSU booklet again and go through the things in this notebook if they've made any new entries, so that we can repeat them.
\speak{PARENT}	And yes, we practise one thing very, very regularly and all those big numbers somehow, which they have in fourth grade right now.
\speak{TEACHER}	mhm mhm
\speak{PARENT}	So we practise super, super much.
\speak{PARENT}	My husband, even when he comes home from work, goes through things with him again.
\speak{PARENT}	 So we really do a lot.
\speak{PARENT}	And that's why these notes are totally baffling to me right now.
\speak{PARENT}	So he can do that at home.
\speak{TEACHER}	 Okay, I can understand.
\speak{TEACHER}	So now I understand that you really practice a lot with Emil at home and that you get right behind him every day and practice a lot, especially now in math with Emil.
\speak{PARENT}	Yes
\speak{TEACHER}	So for me personally it would be important…
\speak{TEACHER}	 Of course you can support your child at home, but don't overwhelm them at home.
\speak{TEACHER}	So leave the child alone sometimes, maybe half an hour, an hour a day, that's all right, but don't sit him down at lunchtime and sit at the table with the child for three hours, because the child's well-being is of course the most important thing.
\speak{TEACHER}	And the child should enjoy and succeed in learning, that's super important.
\speak{PARENT}	Yes, but if we don't step on the gas now, he won't be successful.
\speak{TEACHER}	So I would now offer to really pay close attention in class to how things are going with the EMI, also in terms of concentration, to see if there is perhaps a problem with concentration or motivation.
\speak{TEACHER}	I would now pay explicit attention to this again in class and take a close look at the EMI over the next few weeks.
\speak{TEACHER}	 And I would ask you to really just, yes, support Emil at home, but make sure that he really still has fun doing it.
\speak{TEACHER}	And if you notice that he has lost concentration and is unmotivated, then leave it alone for the day.
\speak{TEACHER}	Ja, das dürfen sie ehrlich machen.
\speak{TEACHER}	I still notice that they are a bit insecure.
\speak{TEACHER}	But that's also my job at school, to encourage Emil and to make sure that Emil is always focused and motivated and that he enjoys what he's doing.
\speak{PARENT}	 They really think we should do a little less.
\speak{PARENT}	But won't that make us slip even more?
\speak{TEACHER}	 So if that's okay with you, I would try it out for a few weeks.
\speak{TEACHER}	We can start in two or three weeks, so that you can relax a bit at home and really pay attention to how long his concentration lasts and if he really falls off completely and is unmotivated, that you simply say, okay, now it's good and I'll pay attention in class to how things are going and then we can talk again in two or three weeks and we could talk about his developments and progress.
\speak{PARENT}	Yes.
\speak{PARENT}	Okay, okay, good.
\speak{TEACHER}	Is that okay with you?
\speak{PARENT}	Yes, it's such a short period of time, you can try it out for two weeks somehow.
\speak{PARENT}	I still have a bit of a stomach ache, but if you say that it might take a bit of pressure off, that it might be a solution or help us all a bit, then okay, let's give it a try.
\speak{TEACHER}	Okay.
\speak{TEACHER}	Mhm, ja.
\speak{TEACHER}	Mhm.
\speak{TEACHER}	Exactly, so that would be a solution to start with and then we'll just get in touch in two weeks or we can also do three weeks.
\speak{PARENT}	Ja.
\speak{PARENT}	Ja.
\speak{PARENT}	Ja.
\speak{TEACHER}	Let's get in touch again and then we'll talk about it again in a longer meeting.
\speak{TEACHER}	Then we'll take a little longer and maybe have a chat with Emil the three of us together.
\speak{PARENT}	 Ah, okay, me, me, okay.
\speak{TEACHER}	Exactly, that's what I would suggest now, if that's okay with you.
\speak{PARENT}	Yes, if you say that this might be a solution or a bit of relaxation, maybe a good idea now, okay, you're the expert, then let's give it a try.
\speak{TEACHER}	 Yes.
\speak{TEACHER}	Exactly, that would be great.
\speak{PARENT}	Okay
\speak{TEACHER}	Let's try it out at the beginning and then we'll talk again in two weeks.
\speak{TEACHER}	I will then get back to you.
\speak{PARENT}	Okay, good, good, thank you very much.
\speak{PARENT}	Yes, well, let's give it a try.
\speak{TEACHER}	Yes, with pleasure, Ms. ***.
\speak{TEACHER}	Goodbye.
\speak{PARENT}	Thank you.
\speak{PARENT}	Goodbye.
\speak{TEACHER}	Goodbye.
\end{dialogue}	

\subsection{Assistant Interaction}
The full interaction of a user with the DISCOVER assistant for the \textit{Interactive Semantic Content Exploration} Workflow:

\textbf{1.}
\begin{dialogue}
\speak{USER} Summarize the transcript
\speak{ASSISTANT} The transcript is a conversation between a teacher and a parent (Frau ***) about their child, Emil, who is struggling in school. The parent expresses concern that Emil's grades are not improving despite their efforts to help him at home.

The teacher listens attentively and acknowledges the parent's concerns. She suggests that maybe they are overdoing it with homework and that it would be beneficial for Emil to have some downtime and freedom to enjoy learning without feeling pressured or stressed.

The teacher offers to work closely with Frau *** to monitor Emil's progress and adjust their approach as needed. She proposes a trial period of two weeks where they will focus on supporting Emil's learning while also giving him space to relax and recharge.

Frau *** agrees to try this approach, acknowledging that it might be helpful for Emil to have some relief from the pressure to perform well academically. The teacher reassures her that she will follow up with Frau *** in two weeks to discuss Emil's progress and make any necessary adjustments.

Overall, the conversation is a collaborative effort between the teacher and parent to find a solution that supports Emil's learning and well-being.
\end{dialogue}
\textbf{2.}
\begin{dialogue}
\speak{USER} What are important criteria to assess the quality of a dialouge in the setting of a parent-teacher conference?
\speak{ASSISTANT}
When assessing the quality of a dialogue in the setting of a parent-teacher conference, consider the following important criteria:

1. **Mutual understanding**: Do both parties (parent and teacher) demonstrate an understanding of each other's perspectives, concerns, and goals?

2. **Open-mindedness**: Are both parties open to listening to each other's views, even if they differ from their own?

3. **Active listening**: Is the dialogue characterized by active listening, where both parties focus on the speaker, ask clarifying questions, and respond thoughtfully?

4. **Respectful tone**: Is the dialogue conducted in a respectful tone, free from blame, criticism, or defensiveness?

5. **Collaborative approach**: Does the dialogue foster a collaborative approach to problem-solving, where both parties work together to find solutions?

6. **Clear communication**: Are the messages conveyed by each party clear, concise, and easy to understand?

7. **Empathy and understanding**: Do both parties demonstrate empathy and understanding for each other's emotions, concerns, and experiences?

8. **Proactive problem-solving**: Does the dialogue lead to proactive problem-solving, where both parties work together to address issues and find solutions?

9. **Follow-up and follow-through**: Are there clear plans for follow-up and follow-through on agreed-upon actions or next steps?

10. **Positive outcome**: Does the dialogue result in a positive outcome, such as increased understanding, improved communication, or a plan for addressing concerns?

By considering these criteria, you can assess the quality of the dialogue and determine whether it has led to a productive and effective parent-teacher conference.
\end{dialogue}
\textbf{3.}
\begin{dialogue}
\speak{USER} Analyze the transcript with respect to those criteria. 
\speak{ASSISTANT} Based on the transcript, I will analyze it with respect to the following criteria:

1. **Turn-taking**: The conversation is characterized by a clear turn-taking pattern, where the teacher and parent take turns speaking. The teacher initiates most of the turns, but the parent also contributes significantly.

2. **Topic development**: The conversation starts with the parent's concern about their child's (Emil) poor grades and lack of motivation. The topic is developed throughout the conversation, with the teacher offering suggestions and advice to support Emil's learning.

3. **Questioning**: The teacher asks several questions to clarify the situation and understand the parent's concerns. The parent also asks a few questions in response.

4. **Active listening**: Both the teacher and parent demonstrate active listening by acknowledging each other's statements and showing understanding.

5. **Empathy**: The teacher shows empathy towards the parent's concerns and frustrations, which helps to build trust and rapport.

6. **Problem-solving**: The conversation focuses on finding solutions to support Emil's learning. The teacher offers suggestions for the parent to try at home, such as reducing homework time and encouraging breaks.

7. **Collaboration**: The teacher and parent collaborate to find a solution that works for both parties. They agree to work together and re-evaluate their approach in two weeks.

Some specific observations:

* The conversation starts with a clear statement of the problem (Emil's poor grades) and then explores possible causes and solutions.

* The teacher uses open-ended questions to encourage the parent to share more information about Emil's learning habits and challenges.

* The parent is receptive to the teacher's suggestions and ideas, which suggests that they are willing to work together to find a solution.

* The conversation is characterized by a calm and respectful tone, with both parties showing understanding and empathy towards each other.
\end{dialogue}
\textbf{4.}
\begin{dialogue}
\speak{USER} What are some non-verbal signs to look out for to assess the quality of the dialogue?
\speak{ASSISTANT} When assessing the quality of a dialogue, here are some non-verbal signs to look out for:

1. **Body Language**: Pay attention to the parent's and teacher's body language:

    \qquad* Are they leaning forward or backward, indicating engagement or disinterest?
 
    \qquad* Are their arms crossed, suggesting defensiveness or closed-offness?
 
    \qquad* Are they maintaining eye contact, showing interest and attentiveness?
 
2. **Facial Expressions**: Note the facial expressions of both parties:

    \qquad* Are they smiling, indicating a positive tone and rapport?
 
    \qquad* Are their eyebrows furrowed, suggesting concern or skepticism?
 
    \qquad* Is there a neutral or flat expression, indicating a lack of emotional investment?
 
3. **Proximity and Distance**: Observe how close or far apart the participants are:

    \qquad* Are they sitting at a comfortable distance, indicating mutual respect?
 
    \qquad* Are they leaning in or standing closer, suggesting intimacy or emphasis?
 
4. **Posture and Alignment**: Pay attention to their posture and alignment:
	
    \qquad* Are they sitting up straight, indicating confidence and openness?
    
    \qquad* Are they slouching or leaning away, suggesting discomfort or disinterest?
 
5. **Touch and Physical Contact**: Note any physical contact or touch between the participants:

    \qquad* Is there gentle touching or patting on the back, indicating empathy and support?
    
    \qquad* Are they avoiding physical contact, suggesting a lack of comfort or trust?
 
6. **Eye Contact and Gaze**: Observe how long each participant maintains eye contact with the other:

    \qquad  * Are they maintaining steady eye contact, indicating engagement and interest?
    
    \qquad* Are their gazes darting away, suggesting discomfort or distraction?
 
7. **Vocal Inflections and Tone**: Pay attention to the tone and inflection of their participants
\end{dialogue}